\documentclass[conference, 10pt]{IEEEtran}
\usepackage[utf8]{inputenc}
\usepackage[utf8]{inputenc}
\usepackage{xcolor}
\usepackage{etoolbox}
\usepackage{upgreek}
\usepackage{amsmath}
\usepackage{comment}
\usepackage{mathtools}
\usepackage[bookmarks=true]{hyperref}
\usepackage{graphicx}
\usepackage{dsfont} %% Enables \mat
\usepackage{comment}
\usepackage{amssymb}
\usepackage{mathtools}
\usepackage{amsthm}
\usepackage{bbm}

\newtheorem{lemma}{Lemma}

\newcommand{\R}{\mathbb{R}}

\newcommand{\N}{\mathbb{N}}

\newcommand{\fcnote}[1]%
    {\textcolor{orange}{\textbf{FC: #1}}}
\newcommand{\twnote}[1]%
    {\textcolor{cyan}{\textbf{TW: #1}}}
\newcommand{\asnote}[1]%
    {\textcolor{blue}{\textbf{AS: #1}}}

\newlength\tindent
\let\sp=\sqparen

\renewcommand{\indent}{\hspace*{\tindent}}
\newcommand{\msnote}[1]%
    {\textcolor{cyan}{\textbf{MS: #1}}}

\newtheorem{assumption}{Assumption}

\newtheorem{theorem}{Theorem}

\title{On the Computational Consequences of Cost Function Design in Nonlinear Optimal Control}
\IEEEoverridecommandlockouts
\author{
    Tyler Westenbroek, Anand Siththaranjan, Mohsin Sarwari, Claire J. Tomlin, and Shankar Sastry$^{1}$ %
    \thanks{$^1$Authors are with the EECS department at the University of California, Berkeley. This work was supported by HICON-LEARN, Defense Advanced Research Projects Agency award number FA8750-18-C-0101,  and the SRC JUMP CONIX Research Center.}
}

\begin{document}
\maketitle
\begin{abstract}
    Optimal control is an essential tool for stabilizing complex nonlinear systems. However, despite the extensive impacts of methods such as receding horizon control, dynamic programming and reinforcement learning, the design of cost functions for a particular system often remains a heuristic-driven process of trial and error. In this paper we seek to gain insights into how the choice of cost function interacts with the underlying structure of the control system and 
    impacts the amount of computation required to obtain a stabilizing controller.
    We treat the cost design problem as a two-step process where the designer specifies outputs for the system that are to be penalized and then modulates the relative weighting of the inputs and the outputs in the cost. To characterize the computational burden associated to obtaining a stabilizing controller with a particular cost, we bound the prediction horizon required by receding horizon methods and the number of iterations required by dynamic programming methods to meet this requirement. Our theoretical results highlight a qualitative separation between what is possible, from a design perspective, when the chosen outputs induce either minimum-phase or non-minimum-phase behavior. Simulation studies indicate that this separation also holds for modern reinforcement learning methods. 
    
\end{abstract}
\section{Introduction}

\label{sec:intro}
The stabilization of complex nonlinear systems is one of the most fundamental and important problems in control theory. Approaches based on optimal control \cite{grimm2005model,bertsekas1996neuro,haarnoja2018soft} form an essential set of tools for solving the stabilization problem and have seen extensive real-world deployment \cite{grizzle2014models,rawlings2017model}. The primary appeal of optimal control is that it allows the user to implicitly encode potentially complex stabilizing controllers as the feedback solutions to certain infinite horizon optimal control problems which are relatively simple to specify.

 In principle, obtaining an optimal infinite horizon controller requires solving the Hamilton-Jacobi-Bellman partial differential equation \cite{bardi1997optimal}. However, for general nonlinear problems it is rarely possible to solve the equation in closed form. This has lead to the development of numerous computational methods which approximate the optimal infinite horizon controller such as dynamic programming \cite{bertsekas1996neuro}, receding horizon control \cite{grimm2005model} and approximate dynamic programming \cite{bertsekas1996neuro} (including modern reinforcement learning methods \cite{haarnoja2018learning}). In one way or another, the parameters of these methods can be used to trade-off the amount of computation that is used with the quality of the resulting approximation.  
 
In this paper we ask: how does the chosen cost function interact with the inherent geometry of the control system to affect the amount of computation needed to obtain a stabilizing controller? To make this question tractable, we consider a two-stage cost design process described below. To concretely characterize the amount of computation needed to obtain a stabilizing controller with a given cost, our theoretical analysis focuses on receding horizon control (RHC) and the dynamic programming algorithm known as value iteration (VI). For these methods we respectively characterize the prediction horizon $T>0$ and number of iterations $k \in N$ the two methods require to stabilize the system. Prior work has shown a direct relationship between these two quantities \cite{bardi1997optimal}, thus we primarily analyze RHC schemes and then use these results to characterize VI. On the empirical side, we investigate how the choice of cost function affects the amount of data modern reinforcement learning algorithms need to stabilize the system. 

The first  step in the cost design process we consider is to select a set of outputs $y=h(x)$ to penalize in the objective, where $x$ is the state of the system. We will then consider running costs of the form $\ell^{\epsilon}(x,u)=\|h(x)\|_2^2 + \epsilon \|u\|_2^2$, where $u$ is the system input and the choice of weighting parameter $\epsilon>0$ represents the second design choice. Intuitively, as ${\epsilon>0}$ is made smaller, the running cost encourages controllers which more aggressively drive the outputs to zero. During our analysis, we will first fix a set of outputs and investigate how the choice of $\epsilon>0$ affects the amount of computation needed to obtain a stabilizing controller.

Our theoretical analysis draws on insights from two distinct lines of work. The first insight comes from the receding horizon literature \cite{grimm2004examples,gaitsgory2015stabilization,relaxing_dp}, which relates the optimal infinite horizon performance for a given cost to the prediction horizon needed by RHC schemes to stabilize the system. Informally, the smaller the infinite horizon cost, the shorter the prediction horizon $T>0$ needed to stabilize the system. The rough intuition here \cite{relaxing_dp} is that as the optimal infinite horizon cost becomes smaller the optimal controller must necessarily drive the running cost to zero more rapidly, and this myopic behavior is easier for RHC schemes to approximate with a short prediction horizon.

The second line of work is the `cheap control' literature \cite{seron1999feedback,aguiar2008performance,saberi1987cheap, sannuti1983direct}, which studies the class of cost functions we consider and draws a separation between minimum-phase (MP) and non-minimum-phase (NMP) systems. In particular, this work bounds the performance of the optimal infinite horizon controller as $\epsilon \to 0$. As the limit is taken the optimal controller drives the outputs to zero as rapidly as possible while ensuring the closed-loop system is asymptotically stable. For MP systems, under suitable conditions, the infinite horizon performance can be made arbitrarily small by taking $\epsilon \to 0$, as a high-gain output-zeroing controller will not destabilize the zeros. However, for NMP systems the unstable zero dynamics present a fundamental barrier to making the infinite horizon performance arbitrarily small, as the optimal controller cannot myopically drive the outputs to zero and must instead `steer' the outputs to stabilize the zeros \cite{seron1999feedback}.

We build on these perspectives by demonstrating that when the chosen outputs lead to MP behavior the prediction horizon $T>0$ needed for RHC schemes to stabilize the system can be made arbitrarily small by making $\epsilon>0$ sufficiently small. Thus, when the system is MP, the user can consistently decrease the computational burden of obtaining a stabilizing controller by using costs which encourage myopically driving the outputs to zero. In sharp contrast, for NMP systems as we take $\epsilon \to 0$ the prediction horizon needed to stabilize the system actually increases and becomes unbounded. Thus, `high-performance' infinite horizon optimal controllers which zero the outputs as rapidly as possible are difficult to approximate in the NMP case but not in the MP case (using RHC and VI).  Moreover, we identify a class of passively unstable NMP systems for which there is a minimum prediction horizon $T>0$ that is needed to stabilize the system \emph{for all choices of} $\epsilon>0$. Taken together, these results demonstrate that NMP dynamics constitute an obstacle, from a computational perspective, to constructing a stabilizing controller. Our simulation studies with reinforcement learning further support this perspective.

\section{Infinite Horizon Optimal Control, Receding Horizon Approximations, And Value Iteration} 
We will consider systems of the form:
\begin{equation}\label{eq:dynamics}
    \dot{x} = f(x) + g(x)u, \ \ \ \ x(0)=x_0, 
\end{equation}
where $x \in \R^n$ is the state, $x_0 \in \R^n$ is the initial condition and $u \in \R^q$ is the input. We will assume that the maps $f \colon \R^n \to \R$ and $g \colon \R^n \to \R^{n\times q}$ are twice continuously differentiable and that $f(0) = 0$. For each $T \in \R\cup \{\infty\}$ we will let $U_T$ denote the set of controls of the form $u\colon \sp{0,T}\to \R^q$ which are measurable and essentially bounded. 
\subsection{Infinite Horizon Optimal Control}
In this section we will consider a generic infinite horizon optimal control problem of the form:
\begin{equation}\label{eq:generic_ocp}
    \inf_{u(\cdot) \in U_\infty} J_\infty(u(\cdot);x_0) = \int_{0}^\infty \ell(x(t),u(t))dt,
\end{equation}
where $(x(\cdot),u(\cdot))$ are subject to \eqref{eq:dynamics} and $\ell \colon \R^n \times \R^q \to \R$ is a twice continuously differentiable running cost which is non-negative, strictly convex in $u$, and satisfies $\ell(0,0)=0$. To the infinite horizon cost we associate the value function:
\begin{equation*}
    V_\infty(x_0) = \inf_{u(\cdot) \in U_\infty} J_\infty(u(\cdot);x_0).
\end{equation*}
We assume that $V_\infty$ is continuous, positive definite, bounded on bounded sets, and that there exists a  control which achieves the optimal performance, namely, $V_\infty(x_0) = J_\infty(u(\cdot);x_0)$ for some $u \in U_\infty$. We will implicitly make these standard assumptions for each of the costs used later in the paper. 

It is well-known that $V_\infty$ can be obtained, in principle, as a solution to the Hamilton-Jacobi-Bellman equation (see e.g. \cite[Chapter 3.2]{bardi1997optimal}) and used to synthesize an optimal stabilizing feedback controller $u_\infty \colon \R^n \to \R^q$ for the cost. This controller is optimal in the sense that when applied to the system it achieves the smallest possible cost from each initial condition. 

\begin{comment}
that the value function satisfies the Hamilton-Jacobi-Bellman (HJB) partial differential equation, nameley,
\begin{equation}
    \inf_{u \in \R^m} \nabla V(x)[f(x)+g(x)u] + \ell(x,u) = 0
\end{equation}
and that an optimal control law which asymptotically stabilizes the system is given by:
\begin{equation}\label{eq:optcontroller}
    u_{\infty}(x) = \arg\min_{u \in \R^m} \nabla V_\infty(x)[f(x) + g(x)u] + \ell(x,u)
\end{equation}
This controller is optimal in the sense that when it is applied to the system \eqref{eq:dynamics} it achieves the minimum performance from each initial condition. If $V_\infty$ is not continuously differentable then an optimal control law can be obtained through the use of viscosity solutions \twnote{reference}. Thus, in principle, the HJB equation can be used to obtain $V_\infty$ and then synthesize the optimal controller with \eqref{eq:optcontroller}. However, as we noted above, it is generally impossible to obtain a closed-form solution for, which has lead to the rise of the approximation methods discussed in the following section. 
\end{comment}
\subsection{Receding Horizon Control}
Next, we discuss receding horizon approximations to the optimal infinite horizon controller $u_\infty$. These schemes use a finite horizon approximation to \eqref{eq:generic_ocp} of the form:
\begin{equation}\label{eq:finite_horizon}
    \inf_{u(\cdot) \in U_T} J_T(u(\cdot);x_0) = \int_{0}^T\ell(x(t),u(t))dt,
\end{equation}
where $T>0$ is a finite prediction horizon. The value function associated to the approximation is:
\begin{equation*}
    V_T(x_0) = \inf_{u(\cdot)\in U_T} J_T(u(\cdot);x_0).
\end{equation*}
To ease exposition we assume that for each $T>0$ and $x_0 \in \R^n$ there exists a unique minimizer $u_{T}(\cdot;x_0)\in U_T$ for the optimal control problem, and we will let $x_{T}(\cdot;x_0)$ denote the corresponding state trajectory. However, we note that in the case where there are multiple optimal controls for a given initial condition the following discussion goes through if any of these control signals are used. 

For each prediction horizon $T >0$ and sampling rate $T\geq \Delta t>0$, receding horizon schemes approximate the infinite horizon controller $u_{\infty}$ with a sampled data control law of the form $u_{T,\Delta t}(t;x_0) = u_{T}(t-t_k;x_{T,\Delta t}(t_k;x_0))$ for each $t\in [t_{k},t_{k+1} )$, where the $t_k = k\Delta t$ are sampling instances and $x_{T,\Delta t}(\cdot;x_0)$ is the state trajectory produced by the receding horizon scheme from the initial condition $x_{0} \in \R^n$. In words, at each sampling instant $t_{k}$ the receding horizon controller solves the finite horizon optimal control problem \eqref{eq:finite_horizon} from the current system state, applies the resulting open loop control for $\Delta t$ seconds, and then repeats the process at time $t_{k+1}$.

At their core, stability results from the literature \cite{jadbabaie2005stability,grimm2005model} are founded on the notion that as $T>0$ increases the RHC scheme more closely approximates the infinite horizon continuous-time feedback controller $u_\infty$, and the quality of this approximation is characterized by how the values of $V_T$ converge to those of $V_\infty$. However, increasing $T$ comes at the cost of additional computational complexity when solving \eqref{eq:finite_horizon}. We next describe a specific stability result used in our analysis, which upper-bounds the prediction horizon $T>0$ needed to ensure stability. In what follows, we will let $\sigma \colon \R^n \to \R$ be a fixed positive definite function which will be used to measure the distance of the state to the origin. 

\begin{assumption}\label{asm:vbound}
There exists $\bar{\alpha}_V>0$ such that:
\begin{equation*}
    V_\infty(x) \leq \bar{\alpha}_V \sigma(x) \ \ \ \ \forall x \in \R^n.
\end{equation*}
\end{assumption}

\begin{assumption}\label{asm:detectability}
There exists a continuously differentiable function $W\colon \R^n \to \R$ and $\bar{\alpha}_W,\underline{\alpha}_W, K_W>0$ such that for each $x \in \R^n$ and $u \in \R^q$:
\begin{equation*}
\underline{\alpha}_W\sigma(x)\leq W(x) \leq \bar{\alpha}_W \sigma(x)
\end{equation*}
\begin{equation*}
\frac{d}{dx}W(x)[f(x) +g(x)u]\leq -K_W \sigma(x) + \ell(x,u).
\end{equation*}
\end{assumption}

 Intuitively, the smaller the constant $\bar{\alpha}_V>0$ in Assumption \ref{asm:vbound} the more rapidly the infinite horizon optimal controller must drive the instantaneous performance to zero. The existence of the map $W$ in Assumption \ref{asm:detectability} ensures that the state measure $\sigma(\cdot)$ is detectable with respect to the loss function $\ell$, in the sense that if $\ell(x,u) = 0$ then we must have $\frac{d}{dt}W(x) <0$ if $x \neq 0$. The following result, which we prove in the supplementary document \cite{tech_rep}, is essentially a continuous-time adaptation and specialization of the stability result from \cite{grimm2005model} which is stated for discrete-time receding horizon control: 

\begin{theorem}\label{thm:mpc}
Let Assumptions \ref{asm:vbound} and  \ref{asm:detectability} hold. Then the receding horizon controller $u_{T,\Delta t}$ will asymptotically stabilize \eqref{eq:dynamics} for each for each $T\geq\Delta t\geq0$ such that:
\begin{equation}
    T>\frac{\bar{\alpha}_V(\bar{\alpha}_V + \bar{\alpha}_W)}{K_W \underline{\alpha}_W (1-M(\Delta t))}
\end{equation}
where 
\begin{equation}
    M(\Delta t) = \exp(-K_W\frac{\underline{\alpha}_W \Delta t}{\bar{\alpha}_V + \bar{\alpha}_W})]< 1.
\end{equation}
\end{theorem}
 Note how the bound on the required prediction horizon $T>0$ depends on the performance of the infinite horizon cost. As $\bar{\alpha}_V$ decreases we can ensure asymptotic stability of the closed-loop system by using RHC schemes with smaller and smaller prediction horizons. This will form the basis for our stability results for minimum-phase systems, when combined with the cheap control results in Section \ref{sec:cheap_control}.

\subsection{Connections Between RHC and VI}\label{sec:VI}
Again using a sampling interval $\Delta t>0$, the value iteration (VI) algorithm constructs a sequence of approximations $ V^1, V_2, \dots$ to the infinite horizon value function $V_\infty$ using the following recursion with $V^{1} (\cdot)\equiv 0$:
\begin{equation}\label{eq:dp}
    V^{k+1}(x_0) = \inf_{u \in U_{\Delta t}} \bigg[\int_{0}^{\Delta t} \ell(x(t),u(t)) dt + V^{k}(x(\Delta t)) \bigg].
\end{equation}
For each $k \in \N$ let $u^k(\cdot;x_0) \in U_{\Delta t}$ be a control which solves the optimization on the right-hand side of \eqref{eq:dp}. Under mild conditions one can show that $V^k$ converges to $V_\infty$ as $k \to \infty$ (see \cite[Chapter 3.3]{bardi1997optimal} for a more in depth discussion). The algorithm produces a sampled-data control law of the form $u^{k,\Delta t}(t;x_0) = u^{k}(t-t_{k};x^{k, \Delta t}(t_k))$ for each $t \in [t_k,t_{k+1})$ where $x^{k,\Delta t}(\cdot;x_0)$ is the state trajectory produced by the control scheme from the given initial condition $x_0 \in \R^n$.  

Using the Principle of Optimality (see e.g. \cite[Prop 3.2]{bardi1997optimal}), one can show that for each $k \in \N$  VI produces the estimate $V^k = V_{k \cdot \Delta t}$. Moreover the $k$-th greedy control is characterized for each $x_0 \in \R^n$ by $u^k(\cdot;x_0) = u_T(\cdot;x_0)|_{\sp{0,\Delta t}}$. Thus, the sampled data controller obtained after $k$ steps of VI with discretization parameter $\Delta t$ is equivalent to the receding horizon controller with prediction horizon $T =k\Delta t$ and re-planning interval $\Delta t$. Thus, the dynamic programming based controller $u^{k,\Delta t}$ implicitly optimizes over the prediction horizon $T =k \Delta t$, and there is a direct correspondence between cases where RHC and VI will stabilize \eqref{eq:dynamics}.

\section{Cost Formulation, Nonlinear Geometry, and the Cheap Control Limit}

\label{sec:background}
We next introduce the cost functions we analyze in this paper. We then highlight how the choice of outputs in the cost interacts with the underlying geometry of the system in Section \ref{sec:geometry}, and then briefly touch on analysis techniques from the cheap control literature in Sections \ref{sec:cheap_control} and \ref{sec:fastslow}. 
\subsection{Cost Formulation}

For each  $\epsilon>0$ and $T>0$ we study the cost functions:
\begin{align}\label{eq:infinite_ocp}
\inf_{u(\cdot) \in U_\infty}J_\infty^\epsilon(u(\cdot);x_0) =\int_0^\infty \|h(x(t))\|_2^2 + \epsilon\|u(t)\|_2^2 dt,
\end{align}
\begin{align}\label{eq:finite_ocp}
\inf_{u(\cdot) \in U_T}J_T^\epsilon(u(\cdot);x_0) =\int_0^T \|h(x(t))\|_2^2 + \epsilon\|u(t)\|_2^2dt,
\end{align}
where $h\colon \R^n \to \R^p$ is a map constructed by the user, which is assumed to be twice continuously differentiable and such that $h(0) =0$. The map $h$ captures the physical quantities of the system which are the most important to drive to zero to meet the control objectives of the designer.

To each of these problems we associate value functions:
\begin{equation}\label{eq:infinite_value}
    V_\infty^{\epsilon}(x_0) = \inf_{u \in U_\infty} J_\infty^\epsilon(u(\cdot); x_0),
\end{equation}
\begin{equation}\label{eq:finite_value}
    V_T^{\epsilon}(x_0) = \inf_{u \in U_T} J_T^\epsilon(u(\cdot); x_0).
\end{equation}
We assume that for each $\epsilon>0$ $V_\infty^{\epsilon}$ is positive definite, continuous and bounded on bounded sets so that the infinite horizon controller will stabilize the system \cite{bardi1997optimal}.

\subsection{Normal Forms, Zero Dynamics and Structural Assumptions}\label{sec:geometry}
To better understand how the cost functions introduced in the previous section interact with the geometry of the state-space model \eqref{eq:dynamics}, let us formally append a set of outputs to the system and form an input-output model of the form: 
\begin{align}\label{eq:io_model}
    \dot{x} &=f(x) + g(x)u \\ \nonumber
    y&=h(x),
\end{align}
where $y \in \R^p$. As alluded to above, the perspective here is that the choice of the running cost \emph{implicitly induces this structure}, and our goal throughout the paper is to understand how this choice impacts the design trade-offs available to the user. Thus, in this section we briefly review basic concepts from nonlinear geometric control which shed light onto how the choice of cost function interacts with the underlying structure of the dynamics. In particular, we discuss how to construct a `normal form' associated to the input-output system \eqref{eq:io_model}, and also introduce several simplifying assumption we will make throughout the paper. Our introduction to these concepts will be brief, as they are covered in many standard references (e.g. \cite[Chapter 9]{sastry2013nonlinear}). We first make the following Assumption:

\begin{assumption}\label{asm:square}
The number of inputs is equal to the number of outputs, namely, $q =p$. 
\end{assumption}

We make this assumption as the construction of normal forms is more straightforward for `square' systems. We discuss the impact of removing this structural condition, along with Assumptions \ref{asm:relative_degree} and \ref{asm:strict_fb} below, in Section \ref{sec:assumptions}.

To obtain a more direct expression relating the evolution of the outputs to the inputs, one can repeatedly differentiate each of the outputs along the dynamics \eqref{eq:dynamics} until an expression of the following form is obtained: 
\begin{equation}
    \begin{bmatrix}
    y_1^{(r_1)} & \dots & y_q^{(r_q)}
    \end{bmatrix}^T = b(x) + A(x)u,
\end{equation}
where $y_j^{(k)}$ denotes the $k$-th time derivative of $y_j =h_{j}(x)$ (the $j$-th entry of the output) and the $r_j$ are positive integers. If the matrix $A(x)$ is bounded away from singularity for each $x \in \R^n$ then we say that the system has a well-defined (vector) relative degree $\bar{r}=(r_1, r_2, \dots, r_q)$. In this case there exists a valid change of coordinates $x \to (\xi,\eta)$  such that in the new coordinates the dynamics are of the form:
\begin{align}\label{eq:normal_form}
    \dot{\xi} &= F\xi + G\bigg[\bar{b}(\xi,\eta) + \bar{A}(\xi,\eta)u \bigg]\\ \nonumber
    \dot{\eta} &= \bar{q}(\xi,\eta) + \bar{P}(\xi,\eta)u
    \\ \nonumber
    y &= C\xi,
\end{align}
where $(F,G)$ is controllable, $(F,C)$ is observable and  $\bar{A}(\xi,\eta)$ is bounded away from singularity for each $(\xi,\eta) \in \R^n$. Here the coordinates $\xi \in \R^{|\bar{r}|}$ capture the outputs and dervatives up to the appropriate order and $\eta \in \R^{n-|\bar{r}|}$ completes the change of coordinates. Namely, $\xi$ contains entries of the form $\xi_{j,k} = y_{j}^{(k-1)}$ for $j = 1, \dots, q$ and $k = 1, \dots, r_j$.  

Next we discuss two simplifying assumption we will use, which are in line with the cheap control literature \cite{braslavsky1998near}:
\begin{assumption}\label{asm:relative_degree}
There exists $r \in \N$ such that the vector relative degree of the system \eqref{eq:dynamics} is $\bar{r}=(r,r, \dots, r)$.
\end{assumption}

Under assumption \ref{asm:relative_degree}, we can arrange $\xi=(\xi_1,\xi_2, \dots{\xi}_r)$ and the $F$, $G$ and $C$ matrices in \eqref{eq:normal_form} to be of the form:
 \begin{equation}\label{eq:matrix_def}
    F = \begin{bmatrix}
        0 & I&  \dots &0\\
      
        \vdots &  & \ddots  &\vdots\\
        0 & 0 & \dots &I\\
        0 & 0 & \dots &0
        \end{bmatrix} \ \ G = \begin{bmatrix}
         0 \\
         \vdots 
         \\
         0\\
         I
        \end{bmatrix},C = \begin{bmatrix}
         I &0 &\dots & 0
        \end{bmatrix}.
\end{equation}
Namely, $\xi_{1} = (y_1,\dots,y_{q})$ represents the outputs and for $k = 2, \dots, r$ the coordinates $\xi_{k} = (y_1^{(k-1)},\dots, y_{q}^{(k-1)})$ capture the $(k-1)-th$ derivatives of the outputs. 

Next, we will restrict the structure of the interconnection between the $\xi$ and $\eta$ subsystems. We say that the input-output system \eqref{eq:dynamics} can be put into \emph{strict feedback form} if the $\eta$ coordinates can be chosen so that the normal form of the dynamics takes the following form:
 \begin{align}\label{eq:strict_fb}
    \dot{\xi} &= F\xi+ G[\bar{b}(\xi,\eta) + \bar{A}(\xi,\eta)u] \\ \nonumber
    \dot{\eta} &= \bar{f}_0(\eta) + \bar{g}_0(\eta)\xi_1\\
    y&= \xi_1 = C\xi, \nonumber
\end{align}
where $\bar{f}_0 \colon \R^{n-|\bar{r}|} \to \R^{n-|\bar{r}|}$ and $\bar{g}_0 \colon \R^{n-|\bar{r}|} \to \R^{(n-|\bar{r}|) \times q}$. 

\begin{assumption}\label{asm:strict_fb}
The input-output system \eqref{eq:dynamics} can be put into the strict feedback form \eqref{eq:strict_fb}.
\end{assumption}

This assumption forbids the input and the derivatives of the outputs to appear directly in the dynamics of the zeros, and prevents the so-called \emph{peaking phenomena} \cite{sepulchre2012constructive}, which we discuss in more detail in Section \ref{sec:assumptions}. In the special case of linear dynamics the system can always be put into strict feedback form when Assumption \ref{asm:relative_degree} is satisfied (see e.g. \cite{sannuti1983direct}). 

For this class of system the \emph{zero dynamics} are obtained by setting the outputs to zero: 
\begin{equation}
    \dot{\eta} = \bar{f}_0(\eta).
\end{equation}
We say that the input-output system is \emph{minimum-phase} (MP) if the zero dynamics are asymptotically stable, and \emph{exponentially minimum-phase} if they are exponentially stable. We say that this system is \emph{non-minimum-phase} (NMP) if it is not minumum-phase, and \emph{exponentially non-minimum-phase} if the dynamics $\dot{\eta} =-\bar{f}_0(\eta)$ are exponentially stable. Finally, if $|\bar{r}| = n$ then no $\eta$ coordinates are needed in the coordinate transformation, and the system is called full-sate linearizable. By way of convention, systems which are full-state linearizable are trivially (exponentially) minimum-phase.

\subsection{The Cheap Control Limit}\label{sec:cheap_control}

The focus of the cheap control literature has been to characterize the structure of the optimal value function $V_\infty^\epsilon$ and corresponding optimal controller $u^\epsilon$ for small values of $\epsilon>0$. In particular, the limiting value $\lim_{\epsilon \to 0}V_\infty^\epsilon(x)$ provides qualitative insight into how difficult it is to drive the chosen outputs to zero from the state $x \in \R^n$ while also stabilizing the internal dynamics. The essential result from the literature is a qualitative separation between the performance limitations of MP and NMP systems. While the majority of the literature has focused on the case where the dynamics are linear \cite{saberi1987cheap,sannuti1983direct,francis1979optimal}, \cite{seron1999feedback} and \cite{braslavsky1998near} extend these results to nonlinear strict-feedback systems of the form \eqref{eq:strict_fb}. An integral part of the analysis for strict feedback systems is the `minimum energy problem' which is formulated using the normal form \eqref{eq:strict_fb}:
\begin{equation}\label{eq:me_problem}
\hat{V}_0(\eta_0) = \inf_{\xi_1(\cdot)} J_0(\xi_1(\cdot); \eta_0)= \int_0^{\infty} \|\xi_1(t) \|_2^2 dt
\end{equation}
where $\dot{\eta} = f_0(\eta) + g_0(\eta)\xi_1$, the output $\xi_1(\cdot)$ is viewed as an `input' to the zero subsystem and the infimum in \eqref{eq:me_problem} is understood to be over $\xi_1(\cdot)$ which drive $\eta(t) \to 0$ asymptotically.  Thus, $\hat{V}_0(\eta)$ can be interpreted as the minimum `energy' of the outputs (in an $\mathcal{L}_2$ sense) that must be accrued by a feedback controller which stabilizes the internal dynamics. When $\hat{V}_0$ is continuously differentiable, an `optimal controller' for the zeros subsystem is given by $\xi_1(t) =\mu_0(\eta(t))$, where $\mu_0 \colon \R^{n-|\bar{r}|} \to \R^q$ is obtain from the HJB PDE associated to the reduced problem \eqref{eq:me_problem}.

Crucially, one may observe that if the system is MP then $\hat{V}_0(\cdot) \equiv 0$ since no `energy' must be expended to stabilize the zeros.  On the other hand, when the system is NMP we will have $\hat{V}_0(\cdot) \not\equiv 0$ since the outputs must be `steered' to stabilize the zeros. In both cases, under suitable technical conditions, the performance limitation for the system is given by  $\lim_{\epsilon \to 0} \hat{V}_\infty^\epsilon(\xi,\eta) = \hat{V}_0(\eta)$, where $\hat{V}_\infty^\epsilon(\xi,\eta)$ is the representation of the value function in the normal coordinates. Thus, for MP systems the infinite horizon cost can be made arbitarily small by taking $\epsilon \to 0$, while there is a fundamental limitation for NMP systems. For MP systems as $\epsilon \to 0$ the optimal controller drives the outputs  directly to zero more and more rapidly, while in the NMP case a high-gain feedback controller drives $\xi(t) \to \mu_{0}(\eta(t))$ to stabilize the zeros \cite{braslavsky1998near}.

\begin{comment}
However, for more general nonlinear systems of the form \eqref{eq:normal_form} it is not clear that even if the system is asymptotically or even exponentially minimum-phase that we will have $\lim_{\epsilon \to 0} V^\epsilon(x) =0$ in the case where $r>1$ due to the well documented \emph{peaking phenomena} (see.\cite{sepulchre2012constructive} for a comprehensive discussion). Essentially, this property dictates that as we increase the gain of a feedback controller which drives $y(t) \to 0$ both the derivatives of the outputs  and the input will `peak' to larger and larger transient magnitudes, potentially perturbing the zeros to a greater and greater extent. Thus, as examples demonstrate \cite{sepulchre2012constructive}, without additional structural assumptions it may be impossible to rule out that a high-gain controller which rapidly drives $y\to 0$ (which is a necessity for $\lim_{\epsilon \to 0} V^{\epsilon}(x)$) does not cause the zeros to escape to infinity. 

\end{comment}

\subsection{Fast-Slow Representations}\label{sec:fastslow}
As mentioned above, singular perturbation techniques play a crucial role in obtaining the aforementioned results and play an essential role in our analysis. Even though most of our arguments are relegated to the supporting document \cite{tech_rep}, it is worthwhile to outline the broad strokes of the analysis here.

In particular, first define the new parameter $\tilde{\epsilon}>0$ such that $\epsilon = \tilde{\epsilon}^{2r}$ so that we may rewrite the running as $\|\xi_1\|_2^2 + \tilde{\epsilon}^{2r}\|u \|_2^2$. If we then define the new coordinates
\begin{equation}\label{eq:transformation}
    \tilde \xi = S(\tilde{\epsilon}) \xi  \ \ \ \ \text{ and } \ \ 
    \tilde u = \tilde{\epsilon}^{r}u,
\end{equation}
where 
\begin{equation}\label{eq:scaling_matrix}
    S(\tilde{\epsilon}) = \text{diag}(1,\tilde{\epsilon},...,\tilde{\epsilon}^{r-1})
\end{equation}
then the system \eqref{eq:strict_fb} takes on the fast-slow representation:
\begin{align}\label{eq:transformed_dyn}
\tilde{\epsilon}\dot{\tilde{\xi}} &= F\tilde{\xi} + G\big[\tilde{\epsilon}^r\tilde{b}(\tilde{\xi}, \eta) +\tilde{A}(\tilde{\xi},\eta)\tilde{u}\big]\\ \nonumber
\dot{\eta} &= \bar{f}_0(\eta) + \bar{g}_0(\eta)\tilde{\xi}_1, 
\end{align}
where $\tilde{b}(\tilde{\xi},\eta) = \bar{b}(S(\tilde{\epsilon})^{-1}\tilde{\xi},\eta)$ and $\tilde{A}(\tilde{\xi},\eta) = \bar{A}(S(\tilde{\epsilon})^{-1}\tilde{\xi},\eta)$ and we have suppressed the dependence of these terms on $\tilde{\epsilon}$. In the re-scaled coordinates the infinite horizon cost becomes: 
\begin{equation}\label{eq:new_cost1}
    \inf_{\tilde{u}(\cdot) \in U_\infty} \tilde{J}_\infty^{\tilde{\epsilon}}(\tilde{u}(\cdot);\tilde{\xi}_0,\eta_0) = \int_{0}^\infty \|\tilde{\xi}_1(t)\|_2^2 + \|\tilde{u}(t)\|_2^2 dt. 
\end{equation}
We will let $\tilde{V}_\infty^{\tilde{\epsilon}}$ denote the representation of the value function in the new coordinates, and we define the reparameterized finite horizon cost $\tilde{J}_T^{\tilde{\epsilon}}$ and optimal performance $\tilde{V}_T^{\tilde{\epsilon}}$ in an analogous way. The form of the re-scaled cost function and dynamics clearly evokes the intuition that we should expect a fast transient for the outputs for small values of $\tilde{\epsilon}>0$.

\section{The Computational Cosequences of Cost Design for Nonlinear Optimal Control}\label{sec:resutlts}
We are now ready to present our theoretical results and draw a qualitative distinction between what is possible, from a computational perspective, when the outputs $y =h(x)$ correspond to either minimum-phase or non-minimum-phase behavior. Due to space constraints, proofs of the following results are relegated to \cite{tech_rep} and we only outline the main arguments here. In Section \ref{sec:performance} we introduce bounds on $V_\infty^\epsilon$ and $V_T^\epsilon$ which are used in the proofs of the main results and provide qualitative insight into how well receding horizon controllers approximate $u_\infty^\epsilon$. 
We then discuss our stability result for minimum-phase systems in Section \ref{sec:mp} and instability results for non-minimum-phase system in Section \ref{sec:nmp}.

\subsection{Performance Bounds}\label{sec:performance}
Our results require the following growth assumptions:

\begin{assumption}\label{asm:growth}
There exists $C>0$ such that the following conditons hold for each $x \in \R^n$ and $(\xi,\eta) \in \R^n:$
\begin{align*} \centering
&\|f(x)\|_2 \leq C  \|x\|_2,  &\|\bar{b}(\xi,\eta) \|_2 \leq C\left(\|\xi\|_2 + \|\eta\|_2 \right),  \\
&\|g(x)\|_2 \leq C, & \|\bar{A}(\xi,\eta)\|_2 < C, \\
&\| \bar{f}_0(\eta)\|_2 \leq C \| \eta\|_2 , &  \|\bar{g}_0(\eta)\|_2 \leq C.
\end{align*}
\end{assumption}

Under this regularity condition we can obtain the following bound on the infinite horizon cost for MP systems:  
\begin{lemma}\label{lemma:growth}
Let Assumptions \ref{asm:square}-\ref{asm:growth} hold. Further assume that \eqref{eq:strict_fb} is exponentially minimum-phase (including full-state linearizable). Then there exist $\hat{K}>0$ such that for each ${0<\tilde{\epsilon} \leq 1}$ we have for each $(\tilde{\xi},\eta)\in \R^n:$
\begin{equation}
\tilde{V}_\infty^{\tilde{\epsilon}}(\tilde{\xi},\eta) \leq \hat{K}\big(\tilde{\epsilon}\|\tilde{\xi}\|_2^2 + \tilde{\epsilon}^{2r} \|\eta\|_2^2\big).
\end{equation}
\end{lemma}
The proof uses the fast-slow representation of the dynamics \eqref{eq:transformed_dyn} and bounds the infinite horizon performance of a sub-optimal feedback linearizing controller of the form $\tilde{u} = \tilde{A}^{-1}(\tilde{\xi},\eta)[-\tilde{\epsilon}^{r}\tilde{b}(\tilde{\xi},\eta) + K\tilde{\xi}]$ where $F+GK$ is Hurwitz, which drives the $\tilde{\xi}$ coordinates to zero exponentially at a rate on the order of $\frac{1}{\tilde{\epsilon}}$. Because the zero dynamics are exponentially minimum-phase, and we have restricted the interconnection between the two systems with Assumption \ref{asm:strict_fb}, this high gain feedback does not destabilize the zeros. As expected, Lemma \ref{lemma:growth} implies that the infinite horizon optimal controller drives the outputs to zero more and more rapidly as $\tilde{\epsilon} \to 0$. In contrast, we recall from Section \ref{sec:cheap_control} that $\tilde{V}_\infty^{\tilde{\epsilon}}$ can be lower-bounded uniformly in the case where the system is NMP, since the outputs must be `steered' so as to stabilize the zeros. 

Next, we discuss a bound on the finite-horizon performance which holds for both MP and NMP systems: 
\begin{lemma}\label{lem:growth_finite}
Let Assumptions \ref{asm:square}-\ref{asm:growth}  hold. Then for each $\bar{T}>0$, there exists $\bar{K}>0$ and $\bar{\epsilon}>0$ such that for each $\bar{T}\geq T > 0$ and $\epsilon \in (0,\bar{\epsilon}]$ we have for each $(\tilde{\xi},\eta) \in \R^n$:
\begin{equation}
\tilde{V}_T^{\tilde{\epsilon}}(\tilde{\xi},\eta) \leq \bar{K}\big(\tilde{\epsilon}\|\tilde{\xi}\|_2^2 + \tilde{\epsilon}^{2r} \|\eta\|_2^2\big).
\end{equation}
\end{lemma}

The result is again obtained by bounding the performance of a linearizing controller which drives the $\tilde{\xi}$ coordinates to zero. Unlike in the infinite horizon case, on bounded time horizons the optimal control does not need to drive the zeros to the origin to achieve a finite cost, which vanishes as $\tilde{\epsilon} \to 0$. Indeed, as the preceding bound indicates, and as we show more formally in the proof of Theorem \ref{thm:nmp} in \cite{tech_rep}, when the prediction horizon $T>0$ is bounded and $\tilde{\epsilon}>0$ is small the optimal control always drives the outputs toward zero in a mypoic fashion. In the NMP case, this will mean that RHC schemes fails to stabilize the zero dynamics when $\tilde{\epsilon}$ is small, unless a very large prediction horizon is used. 

We provided the previous bound in terms of the rescaled coordinates $(\tilde{\xi},\eta)$ and the parameter $\tilde{\epsilon}= \epsilon^{\frac{1}{2r}}$, as doing so cleanly separates how the bound depends on the outputs (and their derivatives) and the zeros. We note that in the original representation the bounds on $V_T^\epsilon(x)$ and $V_\infty^\epsilon(x)$ will both be on the order of $O(\epsilon^{\frac{1}{2r}})$, providing insight into how the relative degree affects the growth of the bound. We state our main results below using the original representation for the problem. 

\subsection{Stability Results and Design Trade-offs for MP Systems}\label{sec:mp}

We are ready to state our main result for MP systems. In the proof of the result we construct a function $W$ which satisfies Assumption \ref{asm:detectability}, and then use the performance bound in Lemma \ref{lemma:growth} to show the following: 
\begin{theorem}\label{thm:full_lin}
Let Assumptions \ref{asm:square}-\ref{asm:growth}  hold. Further assume that \eqref{eq:strict_fb} is exponentially minimum-phase (including full state-linearizable). Then for every fixed $T\geq \Delta t>0$ there exists $\bar{\epsilon}>0$ such that for each $\epsilon \in (0,\bar{\epsilon}]$   the receding horizon controller $u_{T,\Delta t}^\epsilon(\cdot;x_0)$ renders the closed-loop system globally exponentially stable.
\end{theorem}

Thus, in the MP case the designer can consistently decrease the amount of computation needed to obtain a stabilizing RHC controller (as measured by the prediction horizon $T>0$) by decreasing $\epsilon>0$ and encouraging the controller to rapidly drive the outputs to zero. In some applications a fast transient for the outputs may be desirable, and there is no tension between meeting the desired performance objectives and the computational burden of the RHC schemes. In other scenarios the high-gain RHC controllers corresponding to small values of $\epsilon>0$ may cause undesirable effects such as chattering or use too much input to meet design specifications. In either case, when the chosen outputs are MP, the designer retains the freedom to fully explore these design trade-offs.

\begin{comment}
\subsection{Exponentially Minimumphase Systems}
Next we state our result for systems which are exponentially minimumphase but not necessarily fullstate linearizable. 
\begin{theorem}\label{thm:min_phase} Let Assumptions \ref{asm:relative_degree}, \ref{asm:strict_fb} and \ref{asm:growth} hold. Further assume that the zero dynamics $\dot{\eta} = f_0(\eta)$ are globally exponentially stable. Then for each $T\geq\Delta t>0$ there exists $\epsilon_0>0$ such that for each $\epsilon \in (0,\epsilon_0]$ the corresponding receding horizon controller renders the closed-loop system globally exponentially stable. 
\end{theorem}
The proof can again be found in the appendix. The proof again uses the performance bound in Lemma \ref{lemma:growth} and candidate $W$ function in Lemma \ref{lemma:min_phase_rate}. However, due to additional coupling terms which arise between the dynamics of the output subsystem and the zeros, in this case we are only able to guarantee that we can drive the system to a ball around the origin of prescribed radius (as specified by $\delta>0$) starting in any desired operating region (as specified by $R>0$). 
\end{comment}
\subsection{Instability Results, Design Trade-offs and Fundamental Limitations for NMP Systems}\label{sec:nmp}

Our main result for NMP systems in Theorem \ref{thm:nmp} below highlights a class of NMP systems for which there exists a uniform lower-bound on the prediction horizon $T>0$ required to stabilize the system with receding horizon methods which holds for all choices of $\epsilon>0$. We first introduce supportive Lemmas which provide some intuition for this result:

\begin{lemma}\label{lem:small_eps}
Let Assumptions \ref{asm:square}-\ref{asm:growth} hold. Further assume that the system \eqref{eq:io_model} is globally exponentially NMP. Then for each $\bar{T}>0$ there exists $\bar{\epsilon}>0$ such that for each $\epsilon \in (0,\bar{\epsilon}]$ and $\bar{T} \geq T \geq \Delta t>0$ the receding horizon controller $u_{T,\Delta t}^{\epsilon}$ fails to stabilize \eqref{eq:dynamics}. 
\end{lemma}
In sharp contrast to the MP case discussed above, Lemma \ref{lem:small_eps} indicates that in the NMP case as we take $\epsilon \to 0$ the time horizon needed for RHC schemes to stabilize the system actually increases and becomes unbounded. In other words, as $\epsilon \to 0$ and the optimal stabilizing controller $u_\infty^\epsilon$ pushes up against the inherent performance limitations of the system, it becomes more difficult to approximate $u_\infty^\epsilon$ with receding horizon schemes (and thus also the VI method). In extreme cases, an RHC controller formulated with small $\epsilon>0$ and insufficiently large $T>0$ can actually destabilize a passively stable NMP system. To see this, consider the linear system: 
\begin{align}\label{eq:stable_sys}
    \begin{bmatrix}
        \dot{\xi} \\ 
        \dot{\eta}
    \end{bmatrix} = \begin{bmatrix}
        -2 & 1 \\
        -10 &1
    \end{bmatrix} \begin{bmatrix}
        \xi \\
        \eta
    \end{bmatrix} + \begin{bmatrix}
        1\\
        0
    \end{bmatrix} u, \ \ \ y =\xi.
\end{align}
Although the un-driven dynamics are exponentially stable, Lemma \ref{lem:big_eps} predicts that for any $T\geq  \Delta t > 0$ the RHC controller $u_{T,\Delta t}^\epsilon$ will destabilize the system if $\epsilon>0$ is too small. 

While Lemma \ref{lem:small_eps} deals with `small' values of $\epsilon>0$, the following result gives conditions under which `large' values of $\epsilon>0$. For passively stable systems such as \eqref{eq:stable_sys}, when $\epsilon>0$ is large the RHC controller will not exert enough control effort to destabilize the system \cite{kohler2021stability} for any value of $T>0$. However, when the dynamics are passively unstable, `large' values of $\epsilon>0$ prevent the RHC controller from exerting enough control effort to stabilize the system unless $T>0$ is sufficiently large: 

\begin{lemma}\label{lem:big_eps}
Let Assumptions \ref{asm:square}-\ref{asm:growth} hold. Further assume that the dynamics $\dot{x} = -f(x)$ are exponentially stable. Then for each $\bar{\epsilon}>0$ there exists $\bar{T}>0$ such that for each $\epsilon>\bar{\epsilon}$ and $\bar{T}\geq T\geq \Delta t>0$ the receding horizon controller $u_{T,\Delta t}^{\epsilon}$ fails to stabilize \eqref{eq:dynamics}.
\end{lemma}

We combine the preceding results to obtain the following:

\begin{theorem}\label{thm:nmp}
Let Assumptions \ref{asm:square}-\ref{asm:growth} hold. Further assume the additional hypotheses of Lemmas \ref{lem:small_eps} and \ref{lem:big_eps} hold. Then there exists $\bar{T}>0$ such that for each $\epsilon>0$ and $\bar{T}\geq T \geq \Delta t>0$ the receding horizon controller $u_{T,\Delta t}^{\epsilon}$ fails to stabilize \eqref{eq:dynamics}.
\end{theorem}

Thus, unlike in the MP case, NMP dynamics can impose a structural obstacle limiting how small the system designer can make the prediction horizon while ensuring the stability of the closed-loop system. Taken together, the preceding results demonstrate that the presence of NMP dynamics (with respect to the outputs chosen when synthesizing the cost function) limits the capabilities of the designer and restricts the set of design trade-offs that can be exploited.

\subsection{Relaxing Assumptions}\label{sec:assumptions}

Finally we briefly discuss when the technical assumptions made in the paper can be relaxed and when there are obstacles to doing so. First, let us discuss the Assumption \ref{asm:strict_fb}, which stipulates that the system can be put into strict feedback form. For more general nonlinear systems of the form \eqref{eq:normal_form}, driving the outputs to zero with high-gain feedback control may destabilize the zero dynamics, even when the system is exponentially MP, due to the well-documented \emph{peaking phenomena} (see.\cite{sepulchre2012constructive} for a comprehensive discussion). Thus, without additional structural assumptions about the interconnection between the two subsystems we cannot guarantee that the system does not suffer from performance limitations.  

Next, consider Assumption \ref{asm:square}, which stipulates that the number of inputs equals the number of outputs. As long as our other structural assumptions hold, there is little difficulty in extending our results to the case where there are fewer inputs than outputs, so long as the outputs can be decoupled by state feedback. On the other hand, when there are more outputs than inputs the results from \cite{braslavsky2002cheap} indicate that the input-output system will suffer from performance limitations, as the output channels cannot be decoupled by state feedback and driven to zero at arbitrary rates. 

Assumption \ref{asm:relative_degree}, which stipulates that each of the outputs has the same relative degree, is made primarily to streamline analysis. When the outputs have different relative degrees, instead of inducing a fast-slow system of the form \eqref{eq:transformed_dyn}, the cheap control problem induces a singular perturbation problem with multiple time-scale which is much more cumbersome to analyze \cite{sannuti1985multiple}. Nonetheless, we believe an extension of our results to these cases is possible. 
\begin{comment}
\subsection{Input Constraints}\label{sec:inputs}
It has also been noted \cite{perez2002cheap} that, even if the system \eqref{eq:dynamics} is minimumphase, constraints on the inputs of the form $\|u\| <k$ will also lead to performance limitations for the system. In particular, these constraints will limit how quickly we can drive the outputs to zero and will again lead to lower-bounds on the infinite-horizon cost. Thus, in light of the preceding discussion, we should expect that it is computationally more difficult to obtain stabilizing controllers for systems with tight constraints on the inputs. This matches practical experiences and is demonstrated empirically in our experiments. 
\end{comment}

\subsection{Value Iteration and Other Computational Considerations}

In practice, numerical VI algorithms typically use a very small time-step $\Delta t>0$. Using the correspondence between RHC and VI discussed in Section \ref{sec:VI}, in the MP case Theorem \ref{thm:full_lin} indicates that for a fixed $\Delta t$ we can reduce the number of iterations VI requires to produce a stabilize controller by decreasing $\epsilon>0$. In particular, the result demonstrates that there exists $\bar{\epsilon}>0$ sufficiently small such that for each $\epsilon \in (0,\bar{\epsilon}]$ VI will produce a controller which stabilizes the system after only one iteration. On the other hand, Theorem \ref{thm:nmp} indicates that there may be a lower-bound to how many iterations are required to stabilize the system in the NMP case. 

An important direction for future work is to characterize how issues related to numerical discretization affect the qualitative results developed here, where we have studied idealized versions of RHC and VI in which continuous-time optimal control problems are solved as a subroutine. While this has allowed us to clearly characterize when interactions between the cost function and the feedback geometry of the system lead to certain fundamental limitations, the high-gain feedback controllers produced by VI and RHC as we take $\epsilon \to 0$ will lead to numerical stability issues for practical implementations of these methods. For example, in the face of stiff dynamics  grid-based VI methods require a very fine mesh to maintain numerical stability, which increases the computational burden of the method. Thus, broadly speaking, we should expect the limitations of numerical approximations schemes to add an additional layer of computational bottlenecks to the ones considered here.

\section{Numerical Experiments with Reinforcement Learning}
While the preceding theoretical analysis applies only to RHC and VI methods, it is reasonable to conjecture that the trade-offs and fundamental limitations we have identified will appear in one form or another for other methods which seek to approximate infinite horizon optimal controllers. To test this hypothesis, we now investigate how the choice of cost function impacts the number of samples needed by modern reinforcement learning methods to learn a stabilizing controller. These methods are best viewed as noisy approximations to dynamic programming \cite{bertsekas1996neuro}. Specifically, the following experiments use the soft actor-critic algorithm \cite{haarnoja2018soft}, which can be viewed as an approximation to the policy iteration algorithm \cite{bertsekas1996neuro}. Due to space constraints, we only report a small subset of our expirements here, leaving a more extensive evaluation to \cite{tech_rep}. 

\begin{comment}
\textbf{Inverted Pendulum}: 
We first consider the dynamics of an inverted pendulum. The states are $(x_1,x_2)= (\theta,\dot{\theta})$, where $\theta$ is the angle of the arm from vertical. Units have been normalized so that the model is of the form:
\begin{align*}
   \begin{bmatrix}
    \dot{x}_1 \\
    \dot{x}_2
    \end{bmatrix} = \begin{bmatrix}
    x_2 \\ 
    \sin(x_2) + u
    \end{bmatrix},
\end{align*}    
Note that the system is fully-state linearizable with the output $y =x_1$. 
\end{comment}

\textbf{Flexible Link Manipulator}: We consider a model for a flexible link manipulator which can have both MP and NMP outputs. The state is $(x_1,x_2,x_3,x_4) = (\theta_1,\dot{\theta}_1,\theta_2,\dot{\theta}_2)$, where $\theta_1$ is the angle of the arm from vertical and $\theta_2$ is the internal angle of the motor. The dynamics are
\begin{equation*}
     \begin{bmatrix}
    \dot{x}_1 \\
    \dot{x}_2 \\
    \dot{x}_3 \\
    \dot{x}_4
    \end{bmatrix} = \begin{bmatrix}
    x_2 \\
    sin(x_1) + K(x_3 - x_1) - \beta_1 x_2\\
    x_4\\
    K(x_1-x_3) -\beta_2 x_4 +u
    \end{bmatrix},
\end{equation*}
 where $K>1$ is a spring coefficient used to model the flexibility of the joint and $\beta_1,\beta_2 \geq 0$ are friction coefficients. One may observe that if the output $y = x_1$ is chosen then the system is full state lineraizable. However when the output $y =x_3$ is chosen the system has a relative degree of two and the zeros are also two dimensional. In this case a Jacobian linearization at the origin reveals that when the model is friction-less ($\beta_1 = \beta_2 =0$) the system is NMP but when damping is present ($\beta_1,\beta_2>0$) the system is MP. 

\begin{figure}[t]
    \centering
\includegraphics[width=0.45\textwidth]{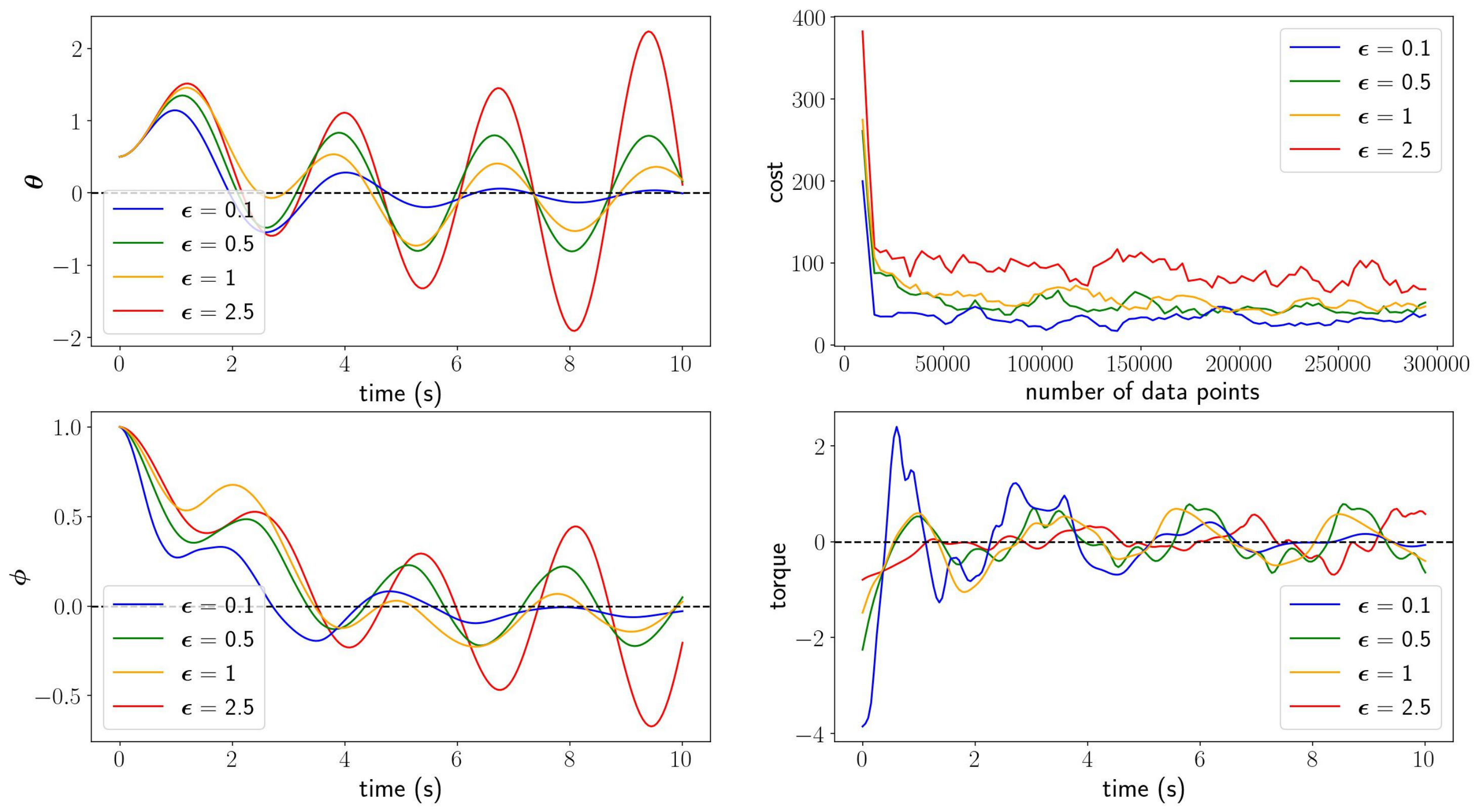}
    \caption{Flexible link manipulator without friction when $y= x_1$.}
    \label{fig:undamped_theta}
    \vspace{-10pt}
\end{figure}

\begin{figure}[t]
    \centering
\includegraphics[width=0.45\textwidth]{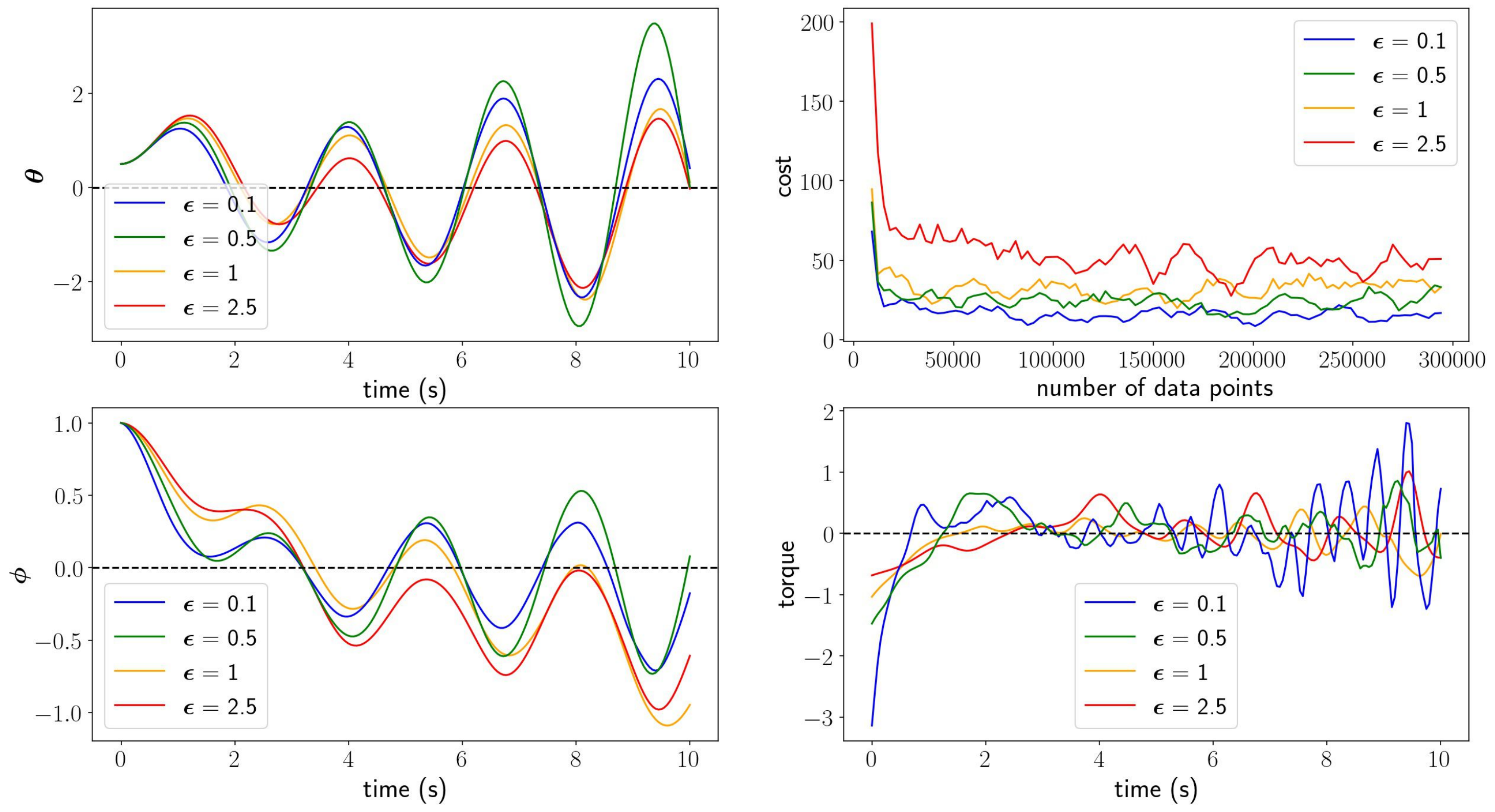}
    \caption{Flexible link manipulator without friction when $y= x_3$.}
    \label{fig:undamped_phi}
    \vspace{-10pt}
\end{figure}

Here, we consider the model without friction, and investigate the damped case in \cite{tech_rep}.  We run experiments for the output $y=x_1$ in Figure \ref{fig:undamped_theta} and the output $y=x_3$ in Figure \ref{fig:undamped_phi}. In each figure we plot the results obtained by training a controller with the soft actor-critic algorithm using different values of the weighting parameter $\epsilon>0$. The upper-right plot in each figure depicts the average cost obtained by the algorithm after obtaining access to different numbers of samples. While the left-most plots depict a trajectory generated by the best-performing controller that was obtained after 300,000 samples of the dynamics. As the figures clearly show, the reinforcement learning algorithm struggles to learn a stabilizing controller when $y=x_3$ for all values of $\epsilon>0$ that were tested. However when $y=x_1$ the algorithm is able to rapidly learn a stabilizing controller for small values of $\epsilon>0$ but again struggles when the parameter is large. Thus, we conclude that the flat output $y = x_1$ is a `better' choice of output, and observe that in these preliminary investigations the trade-offs and limitations we characterized above appear to hold when reinforcement learning algorithms are employed. Full details of the experiments including hyper-parameters, initial state distributions, etc. can be found in \cite{tech_rep}.

\section{Conclusion}
In this paper, we studied how the geometry of a control system introduces computational limitations when practically solving for optimal controllers. Through the lenses of receding horizon control and cheap control, we identify a separation in qualitative behaviour between MP and NMP systems.  Further we experimentally verified these separations when a modern RL algorithm is used as the controller-synthesis tool. 

For future work, we also hope to characterize formally the sample complexity of various reinforcement learning methods when different cost functions are used. The primary challenge for this program is the lack of rigorous convergence proofs for the most common reinforcement learning methods. Indeed, state-of-the art sample complexity guarantees are generally restricted to the case of linear dynamics and require access to an initial stabilizing controller \cite{fazel2018global, krauth2019finite}, yet experience has shown that variations of these methods are none-the-less able to synthesize stabilizing controllers from `scratch'. 

\bibliographystyle{IEEEtran}
\bibliography{IEEEabrv,ref.bib}

%\appendix

%\input{appendix}
%\input{appendix_arxiv}

\end{document}